# A theory for the effect of patch / non-patch attractions on the self-assembly of patchy colloids

Bennett D. Marshall

*ExxonMobil Research and Engineering, 22777 Springwoods Village Parkway, Spring TX 77389 USA*

## Abstract

In this paper, we develop a thermodynamic perturbation theory to describe the self-assembly of patchy colloids which exhibit both patch-patch attractions as well as patch / non-patch attractions. That is, patches attract other patches as well as the no patch region (we call this region $\Psi$). In general, the patch-patch and patch-$\Psi$ attractions operate on different energy scales allowing for a competition between different modes of attraction. This competition may result in anomalous thermodynamic properties. As an application, we tune the patch parameters to reproduce the liquid density (suitably scaled) maximum of water. It is then shown that the liquid branch of the colloids phase diagram has liquid densities consistent with both saturated and super-cooled liquid water. Finally, it is shown that the colloids reproduce water's anomalous minimum in isothermal compressibility and negative volume expansivity.

Email: bennettd1980@gmail.com

# I: Introduction

Self-assembly of patchy colloids has been the subject of extensive experimental[1–4] and theoretical[5–8] investigation. Patchy colloids (PC's) are spherical colloids with discrete surface patches which attract other patches, but have only repulsion between the patches and the non-patch region (the surface area not covered by patches). These patch-patch attractions are highly directional. This anisotropy results in controlled self-assembly into predesigned structures as well as rich phase behavior. Another class of anisotropic colloids are inverse patchy colloids (IPC's)[9–13]. As the name suggest, IPC's are the inverse of PC's. There are some number of surface patches on the colloid, but patch-patch interactions are repulsive instead of attractive, and patch / non-patch attractions are attractive instead of repulsive. It has been demonstrated that IPC's exhibit a rich variety of phase behavior.

What has not been studied (to the authors knowledge) is the class of colloids which exhibit both patch-patch and patch-non-patch attractions. We will call the surface area of the colloid which is not a patch $\Psi$, and will call this class of colloids, competitive colloids. See Fig. 1 for an illustration. An interesting feature of competitive colloids is that certain choices of patch size and attractive energies could set up a competition between patch-patch and patch-$\Psi$ attractions resulting in exotic phase behavior such as a reentrant phase diagram.

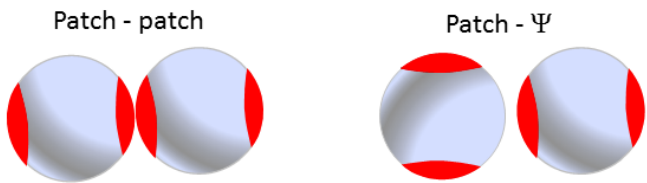

**Figure 1:** Diagram of interacting competitive colloids. The grey region is called $\Psi$. Both patch-patch and patch-$\Psi$ attractions are allowed

Theoretical and Monte Carlo simulation calculations have shown that certain classes of PC's exhibit this type of phase behavior.[14,15] One drawback of these PC models is that the patchy colloids often require exotic patch designs. For instance, the 2A-9B colloids[14] which exhibit a reentrant phase diagram are composed of two large *A* type patches in an axisymmetric location, and 9 smaller *B* type patches along the equator. While well-defined from a theoretical stand point, synthesis of these would be quite challenging. Competitive colloids on the other hand may be comparatively easy to synthesize. It has been experimentally demonstrated that patchy colloids with several patches can be synthesized.[2] The patch-patch attractions are mediated by DNA sticky ends. For the case of competitive colloids, a judicious choice of patch DNA sticky ends and the DNA sequence of Ψ sticky ends would allow researchers to control the patch-patch binding energies as well as the patch-Ψ binding energies.

Wertheim's[16,17] thermodynamic perturbation theory (TPT) for associating fluids has proven to be a valuable tool in the theoretical study of the self-assembly and phase behavior of PC's.[6,14,18,19] Wertheim's reformulation of statistical mechanics into a multi-density formalism is designed to accommodate the strength and limited valence of association interactions. While formally exact in its most general form, when applied as a perturbation several assumptions are made. Perturbation theory is defined by the neglect of all contributions to the change in free energy due to association which represent interactions between distinct associated clusters. This allows the free energy to be described by reference system correlations. When applied at first order (TPT1) it is assumed that all patches are independent of one another (no patch-patch correlations) and all patches are singly bondable. Going to higher order in perturbation allows the inclusion of steric effects between patches[20], rings[20,21] and multiple bonding of patches[22,23]. In addition to perturbation theory, the more complex formalism of integral equation theory (IET) can be employed.[24,25]

Given the remarkable success in applying TPT1 to describe the self-assembly and phase behavior of patchy colloid fluids. It would be equally valuable to have such a tool to describe the self-assembly and phase behavior of competitive colloids. Currently, a TPT has not been developed which could be applied to competitive colloid interaction.

The first challenge in the development of such a theory is the accounting for steric effects between colloids which are simultaneously bonded to Ψ on the same colloid. Previous perturbation theories[19,22,26–28] have been developed to account for multiple bonding of patches. However, applicable forms of these theories have only been developed for patches which bond up to twice. The Ψ region on competitive colloids can receive as many association bonds as sterically allowed. This will, in general, allow for Ψ to receive more than two association bonds. A similar problem was addressed in the development of a TPT for binary mixture of patchy and spherically symmetric colloids where the patches on the patchy colloid can bond to other patches or with the spherically symmetric colloid.[18,29,30] In the development of the TPT for competitive colloids, we treat steric hindrance for $\Psi$ in a similar fashion to the spherically symmetric colloid in these binary mixtures.

With the multiple bonding of $\Psi$ addressed, the challenge is then to develop a general thermodynamic equation of state which can be used for competitive colloids with any number and functionality of singly bonded patches. It is the generality of the TPT1 solution which has led to it's widespread application. Here, we seek a similar universal solution for competitive colloids.

The resulting free energy presented in section II is a very general solution to TPT for the case that there are an arbitrary number and functionality of singly bondable patches which can associate with a large irregular $\Psi$ of arbitrary shape. The number of association bonds which $\Psi$ can receive is dictated by the geometry of $\Psi$ and the resulting steric effects between colloids bonded to $\Psi$. Competitive colloids are one application of this general solution.

The thermodynamics of liquid and super-cooled water has been the subject of intense research for decades.[31] The fascination with water stems from its anomalous thermodynamic properties which result from the structural transition to tetrahedral coordination which accompanies water becoming fully hydrogen bonded.[32] There has also been significant interest in developing patchy colloid models which reproduce these anomalous features of water. Sciortino and coworkers[33,34] have extensively studied 4 patch colloidal models with tetrahedral symmetry using molecular simulation.

Unfortunately, since perturbation theory assumes an unchanging hard sphere reference fluid, it is not possible to reproduce these anomalous properties of water using TPT for colloidal models based solely on the transition to tetrahedral symmetry. That is, in TPT the theory knows that a tetrahedral patchy colloid can only be bonded to 4 other colloids; however, the theory has no knowledge that the patches are arranged in a tetrahedral arrangement. Smallenburg et al.[35] showed that if one considers non-additive hard sphere diameters that TPT1 could be used to predict anomalous features of liquid water. However, as discussed above, it is the tetrahedral symmetry of fully bonded liquid water which results in the unique properties of water, not non-additive hard sphere diameters.

While TPT cannot strictly be used to predict the structural anomalies of water due to tetrahedral symmetry, it is possible to use TPT as a tool to design patchy colloid models which mimic the anomalous behavior of water by setting up a competition of energy scales of attraction.

Recently, Rovigatti *et al.*[36] showed that patchy colloid models which exhibit re-entrant phase behavior (2A-9B colloids[14]) could be used to explain the possible divergence of the compressibility and specific heat in supercooled water.

As a first application of the competitive colloid model, we map the thermodynamic behavior of water onto a competitive colloid potential of interaction. Going beyond qualitative comparison to provide an experimentally testable prediction, we fit competitive colloid parameters (patch size and attractive energies) which reproduce the density of water (suitably scaled) in the temperature range $T = 0$ - $100$ °C. With the parameters fixed, the full phase diagram is predicted. We demonstrate how the density maximum of water can be reproduced using a competitive colloid model. At the point of the density maximum, patch-$\Psi$ association bonds are being traded for patch-patch association bonds. It is also shown that, like liquid water, this results in a minimum in the isothermal compressibility as a function of temperature and negative volume expansivity.

## II: Theory

### A) Potential of interaction

In this section, we describe a simple geometric form for the potential of interaction of spherical colloids of diameter $d$ with anisotropic attractions. We allow for the colloids to have a set of singly bondable attractive patches $\Gamma_P = \{A, B, C,…, Z\}$. These patches can attract each other (or not) as well as have attractions to the surface of the colloid which is not covered by patches. We call this no patch region $\Psi$. The total set of association sites is then $\Gamma = \{\Psi, \Gamma_P\}$.

In TPT1 the size of all patches is restricted such that steric effects prevent a patch bonding more than once. For the patches in $\Gamma_P$ we enforce this single bonding condition. However, in general, the $\Psi$ region may receive several association bonds due to its larger size. For this reason, we do not restrict the interaction between the association sites and $\Psi$. The maximum number of association bonds of $\Psi$ is $n^{max}$ which is set by steric repulsions. This is illustrated in Fig. 2.

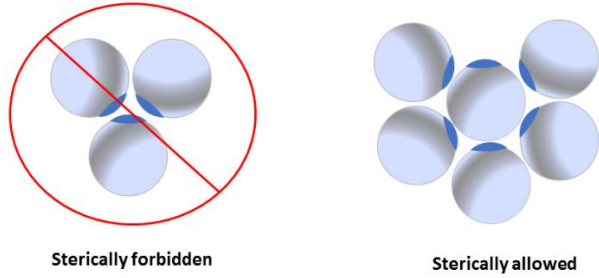

**Figure 2:** Diagram depicting steric restrictions in patch-patch and patch-$\Psi$ interactions

The potential of interactions is then given as the sum of hard sphere + association contributions

$$\varphi(12) = \varphi_{HS}(r_{12}) + \sum_{A \in \Gamma_p} \sum_{B \in \Gamma_p} \varphi_{AB}(12) + \sum_{C \in \Gamma_p} (\varphi_{C\Psi}(12) + \varphi_{\Psi C}(12)) \quad (1)$$

The term $\varphi_{HS}(r)$ is the potential of a hard sphere fluid

$$\varphi_{HS}(r) = \begin{cases} \infty & r < d \\ 0 & Otherwise \end{cases} \quad (2)$$

In Eq. (1) $\varphi_{AB}(12)$ represents the association potential between patch $A$ on colloid 1 and patch $B$ on colloid and $\varphi_{C\Psi}(12)$ represents the association potential between patch C on colloid 1 and non-patch surface $\Psi$ on colloid 2. The notation $(1) \equiv (\vec{r}_1, \Omega_1)$ represents the position $\vec{r}_1$ and orientation $\Omega_1$ of colloid 1 and $r_{12}$ is the distance between centers of the pair.

As a primitive model for the patch-patch potential we employ conical square well association sites(also known as the Kern-Frenkel model of PC's)[37,38]

$$\varphi_{AB}(12) = \begin{cases} -\varepsilon_{AB} & r_{12} \leq r_c \text{ and } \theta_{A1} \leq \theta_{c,A} \text{ and } \theta_{B2} \leq \theta_{c,B} \\ 0 & otherwise \end{cases} \quad (3)$$

Where $r_c$ is the maximum distance between colloids for which association can occur, $\theta_{A1}$ is the angle between the center of patch $A$ on colloid 1 and the vector connecting the two centers, and $\theta_c$ is the maximum angle for which association can occur. With this, if two colloids are both positioned and oriented correctly, a bond is formed and the energy of the system is decreased by a factor $\varepsilon_{AB}$. Here we assume a constant value of $r_c$, but allow $\theta_c$ to vary as long as it is small enough to ensure the single bond per patch condition.

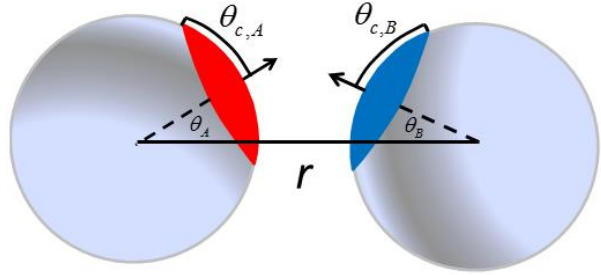

**Figure 3:** Diagram of conical square well patch-patch attraction. The patch size is controlled the critical angle $\theta_c$

Similarly, the patch-$\Psi$ potential is given as the orientationally dependent square well contribution

$$\varphi_{P\Psi}(12) = \begin{cases} -\varepsilon_{P\Psi} & r_{12} \leq r_c \text{ and } \theta_{P1} < \theta_{c,P} \text{ and } \theta_{D2} > \theta_{c,D} \ \forall \ D \in \Gamma_P \\ 0 & \text{otherwise} \end{cases} \quad (4)$$

Equation (4) states that there is a square well attraction of depth $\varepsilon_{P\Psi}$ between the patch $P$ on colloid 1 and the non-patch surface $\Psi$ on colloid 2 if the colloids centers are within a distance $r_c$, the orientations are such that the patch $P$ of colloid 1 is correctly oriented for association $\theta_{P1} < \theta_{c,P}$, but all patches $D$ of colloid 2 are not correctly oriented for association $\theta_{D2} > \theta_{c,D} \ \forall \ D \in \Gamma_p$.

Figure 3 illustrates the interaction of two colloids. If $r < r_c$ with $\theta_A < \theta_{c,A}$ and $\theta_B < \theta_{c,B}$ a patch-patch association bond is formed. However, if $r < r_c$ with $\theta_A < \theta_{c,A}$ and $\theta_B > \theta_{c,B}$ a patch-$\Psi$ bond is formed. The potential Eqns. (1) – (4) is general enough to allow for the inverse patchy colloid type of attraction as well as allow one to probe the effects of patch-patch attractions and patch size on the phase behavior of this class of colloids. We will call this class of colloids competitive colloids. In what follows we apply thermodynamic perturbation theory to develop an analytical model to describe competitive colloids.

**B) Thermodynamic perturbation theory**

We develop the theory in the multi-density formalism of Wertheim[17,39,16] where each bonding state of a colloid is assigned a number density. The density of colloids bonded at the set of sites $\alpha$ is given by $\rho_\alpha$. To aid in the reduction to irreducible graphs, Wertheim introduced the density parameters $\sigma_\gamma$

$$\sigma_\gamma = \sum_{\alpha \subset \gamma} \rho_\alpha \tag{5}$$

where the empty set $\alpha = \varnothing$ is included in the sum. Two notable cases of Eq. (5) are $\sigma_\Gamma = \rho$ and $\sigma_o = \rho_o$; where $\rho$ is the total number density of colloids and $\rho_o$ is the density of colloids not bonded at any patch including $\Psi$ (monomer density).

In Wertheim's multi-density formalism, the exact change in free energy due to association is given by

$$\frac{A - A_{HS}}{Vk_B T} = \frac{A^{AS}}{Vk_B T} = \rho \ln\left(\frac{\rho_o}{\rho}\right) + Q + \rho - \Delta c^{(o)}/V \tag{6}$$

where $V$ is the system volume, $k_B$ is Boltzmann's constant, $T$ is temperature, $A_{HS}$ is the hard sphere reference free energy and $Q$ is given by

$$Q = -\rho + \sum_{\substack{\gamma \subset \Gamma \\ \gamma \neq \varnothing}} c_\gamma \sigma_{\Gamma - \gamma} \tag{7}$$

The term $\Delta c^{(o)}$ is the associative contribution to the fundamental graph sum which encodes all association attractions between the colloids, and $c_\gamma$ is obtained from the relation

$$c_\gamma = \frac{\partial \Delta c^{(o)}/V}{\partial \sigma_{\Gamma - \gamma}} \tag{8}$$

where in Eq. (8) $\gamma \neq \varnothing$. The graph sum $\Delta c^{(o)}$ is decomposed as

$$\Delta c^{(o)} = \Delta c^{(o)}_{pp} + \Delta c^{(o)}_{p\Psi} \tag{9}$$

where $\Delta c^{(o)}_{pp}$ accounts for the attractions between patches in $\Gamma_p$ and $\Delta c^{(o)}_{p\Psi}$ accounts for attractions between patches in $\Gamma_p$ and $\Psi$. We treat the interaction between $\Gamma_p$ patches in first order perturbation theory (TPT1) giving $\Delta c^{(o)}_{pp}$ as[40]

$$\Delta c_{pp}^{(o)}/V = \frac{\xi}{2}\sum_{A\in\Gamma_P}\sum_{B\in\Gamma_P}\sigma_{\Gamma-A}\kappa_{AB}f_{AB}\sigma_{\Gamma-B} \tag{10}$$

In Eq. (10) $\kappa_{AB}=(1-\cos\theta_{c,A})(1-\cos\theta_{c,B})/4$ is the probability that two colloids are oriented such that patch $A$ on colloid 1 can bond to patch $B$ on colloid 2. Also,

$$\xi = 4\pi\int_{d}^{r_c} r^2 y_{HS}(r)dr \tag{11}$$

is the integral of the hard sphere reference cavity correlation function over the bond volume. For this quantity we employ the method of Marshall and Chapman.[20] Lastly, the term $f_{AB} = \exp(\varepsilon_{AB}/k_bT) - 1$ is the magnitude of the association Mayer function.

For the attractions between $\Gamma_p$ patches and $\Psi$ we must explicitly account for the fact that the $\Psi$ (non-patch) region of the colloid can bond up to a maximum of $n^{max}$ times. To accomplish this, we take a similar approach to that taken for mixtures of patchy and spherically symmetric colloids[18,29]. The general form of the graph sum for this interaction remains unchanged (equations (13) – (14) of ref[18]), except for the change that the current approach is for a pure component fluid which results in the change $\rho_o^{(s)} \to \sigma_{\Gamma-\Psi} = \sigma_{\Gamma_p}$ to obtain

$$\frac{\Delta c_{p\Psi}^{(o)}}{V} = \sigma_{\Gamma-\Psi}\sum_{n=1}^{n^{max}}\frac{1}{n!}\Delta^n \delta^{(n)}\Xi^{(n)} \tag{12}$$

In equation (12) $n$ is the number of incident association bonds on $\Psi$ and $\Delta$ is

$$\Delta = y_{HS}(d)\sum_{L\in\Gamma_p}\sigma_{\Gamma-L}f_{L\Psi}\sqrt{\kappa_{LL}} \tag{13}$$

$\delta^{(n)}$ is the second order correction to the first order superposition of the many body correlation function for the associated cluster. This term is evaluated using the branched TPT2 solution of Marshall and Chapman[41] as

$$\delta^{(n)} = \begin{cases} (1+4\lambda)^{\frac{n-3}{2}}\left(\frac{1+\sqrt{1+4\lambda}}{2}\right)^3 & \text{for } n>1 \\ 1 & \text{for } n=1 \end{cases} \tag{14}$$

The term $\lambda = 0.2336\eta + 0.1067\eta^2$ where $\eta$ is the packing fraction.[42] Finally, the terms $\Xi^{(n)}$ are the cluster partition functions which are independent of temperature and density. These terms encode the steric effects rich arise when multiple colloids attempt to donate an association bond to $\Psi$. We will discuss the cluster partition functions in section C.

Now that $\Delta c^{(o)}$ has been fully specified, the densities of the various bonding states can be calculated through the relation[27]

$$\frac{\rho_\gamma}{\rho_o} = \sum_{P(\gamma)=\{\tau\}} \prod_\tau c_\tau \tag{15}$$

where $P(\gamma)$ is the partition of the set $\gamma$ into non-empty subsets. For example, the density $\rho_{ABC}$ is given by $\rho_{ABC} = \rho_o(c_{ABC} + c_{AB}c_C + c_{BC}c_A + c_{CA}c_B + c_A c_B c_C)$. We note the following relation holds

$$c_\gamma = 0 \quad for \quad n(\gamma) > 1 \tag{16}$$

This allows Eq. (15) to be simplified to

$$\frac{\rho_\gamma}{\rho_o} = \prod_{A\in\gamma} \frac{\rho_A}{\rho_o} \tag{17}$$

The relation in (17) implies[40] the following relations for the fraction of colloids *not bonded* at patch $D$

$$X_D = \frac{\sigma_{\Gamma-D}}{\rho} = \frac{1}{1+c_D} \tag{18}$$

as well as the relation for the monomer fraction

$$X_o = \prod_{A\in\Gamma} X_D \tag{19}$$

The term $c_D$ is obtained through (8). For the case $D \in \Gamma_p$

$$c_D = \xi\rho \sum_{A\in\Gamma_P} \kappa_{AD} f_{AD} X_A + \rho X_\Psi y_{HS}(d) \sum_{n=1}^{n^{max}} \frac{n}{n!} \sqrt{\kappa_{DD}} f_{D\Psi} \Delta^{n-1} \Xi^{(n)} \delta^{(n)} \tag{20}$$

and for $D = \Psi$

$$c_\Psi = \sum_{n=1}^{n^{max}} \frac{1}{n!} \Delta^n \delta^{(n)} \Xi^{(n)} = \frac{\Delta c_{p\Psi}^{(o)}/V}{\sigma_{\Gamma_p}} \tag{21}$$

We now define the fraction of colloids with $\Psi$ bonded $n$ times (irrespective of bonding of the patches in $\Gamma_p$) as $\chi_{\Psi,n}$. These fractions must sum to unity

$$\sum_{n=0}^{n^{max}} \chi_{\Psi,n} = 1 \tag{22}$$

Note that $\chi_{\Psi,o} = X_\Psi$ is simply the fraction of colloids not bonded at $\Psi$. Comparing Eqns. (18) and (21) – (22) we deduce the fractions for $n > 0$

$$\chi_{\Psi,n} = \frac{1}{n!} \chi_{\Psi,o} \Delta^n \delta^{(n)} \Xi^{(n)} \qquad for\ n > 0 \tag{23}$$

From Eqn. (7)

$$\frac{Q}{\rho} = -1 + \sum_{A \in \Gamma} c_A X_A \tag{24}$$

which when combined with Eq. (18) gives

$$\frac{Q}{\rho} = -1 + \sum_{A \in \Gamma} (1 - X_A) \tag{25}$$

The patch-patch contribution to the graph sum is found to be

$$\frac{\Delta c_{pp}^{(o)}}{V} = \frac{\rho}{2} \sum_{A \in \Gamma_p} X_A c_A - \frac{\rho}{2} \sum_{n=1}^{n^{max}} n \chi_{\Psi,n} = \frac{\rho}{2} \sum_{A \in \Gamma_p} (1 - X_A) - \frac{\rho}{2} \bar{n}_\Psi \tag{26}$$

where $\bar{n}_\Psi$ is the average number of incident association bonds on $\Psi$

$$\bar{n}_\Psi = \sum_{n=0}^{n^{max}} n \chi_{\Psi,n} \tag{27}$$

From Eqns. (12), (22) - (23)

$$\frac{\Delta c_{p\Psi}^{(o)}}{V} = \rho \sum_{n=1}^{n^{max}} \chi_{\Psi,n} = \rho(1 - \chi_{\Psi,o}) = \rho(1 - X_\Psi) \tag{28}$$

Combining these results, we obtain the association contribution to the free energy ($N$ - total number of colloids)

$$\frac{A^{AS}}{Nk_BT} = \sum_{A \in \Gamma_p}\left(\ln X_A - \frac{X_A}{2} + \frac{1}{2}\right) + \ln X_\Psi + \frac{\bar{n}_\Psi}{2} \tag{29}$$

The contribution within the summation of Eq. (29) gives the standard TPT1 free energy for the patches in $\Gamma_p$, while the remaining terms represent corrections for the fact that $\Psi$ can receive multiple association bonds.

Equation (29) is a very general solution to TPT for the case that there are an arbitrary number and functionality of singly bondable patches in $\Gamma_p$ which can associate with a large irregular $\Psi$ of arbitrary shape and functionality which can receive as many association bonds as are sterically allowed. This multi-body steric hindrance is encoded in the single cluster partition function $\Xi^{(n)}$. While these steric effects are included for multiple bonding of $\Psi$, correlations between separate patches are neglected and treated in first order perturbation theory. This approach allows for a relatively simple and elegant solution, while including full steric effects on the multiple bonding of $\Psi$.

The chemical potential is obtained using the general relation[16]

$$\frac{\mu_{AS}}{k_bT} = \frac{\mu - \mu_{HS}}{k_bT} = \ln X_o - \frac{\partial}{\partial \rho}\frac{\Delta c^{(o)}}{V} \tag{30}$$

Taking the derivative of the individual contributions

$$\frac{\partial}{\partial \rho}\frac{\Delta c_{pp}}{V} = \frac{\rho}{2}\left(\sum_{A \in \Gamma_p}(1 - X_A) - \bar{n}_\Psi\right)\frac{\partial \ln \xi}{\partial \rho} \tag{31}$$

and

$$\frac{\partial}{\partial \rho}\frac{\Delta c_{p\Psi}}{V} = \rho\frac{\partial \ln y_{HS}(d)}{\partial \rho}\sum_{n=1}^{n^{max}}\frac{n}{n!}\chi_{\Psi,o}\Delta^n\delta^{(n)}\Xi^{(n)} + \rho\sum_{n=1}^{n^{max}}\frac{1}{n!}\chi_{\Psi,o}\Delta^n\delta^{(n)}\Xi^{(n)}\frac{\partial \ln \delta^{(n)}}{\partial \rho} \tag{32}$$

$$= \rho\frac{\partial \ln y_{HS}(d)}{\partial \rho}\bar{n}_\Psi + \rho\sum_{n=1}^{n^{max}}\chi_{\Psi,n}\frac{\partial \ln \delta^{(n)}}{\partial \rho}$$

Combining (30) – (32)

$$\frac{\mu_{AS}}{k_b T} = \sum_{A \in \Gamma} \ln X_A - \frac{\rho}{2} \left( \sum_{A \in \Gamma_p} (1 - X_A) - \bar{n}_\Psi \right) \frac{\partial \ln \xi}{\partial \rho} - \rho \left( \frac{\partial \ln y_{HS}(d)}{\partial \rho} \bar{n}_\Psi + \sum_{n=1}^{n^{max}} \chi_{\Psi,n} \frac{\partial \ln \delta^{(n)}}{\partial \rho} \right) \tag{33}$$

The bonding state of the colloid is defined through the fractions $X_A$ from which the fractions $\chi_{\Psi,n}$ can be evaluated. For a colloid with $n_p$ patches in $\Gamma_p$, this in general requires the simultaneous solution of $n_p + 1$ equations defined by Eq. (18) with the quantities $c_D$ defined by equations (20) – (21).

### C) Cluster partition functions

The perturbation theory derived in section B is a general result. Specification to the specific competitive colloid potential defined through Eqns. (1) – (4) is achieved through evaluation of the cluster partition functions $\Xi^{(n)}$. Here $n$ is the number of colloids with patches bonded to $\Psi$ on a center colloid in the cluster. The cluster partition functions are related to the number of states for which a given associated cluster can exist. To derive these functions, we modify the approach previously developed to calculate $\Xi^{(n)}$ for mixtures of patchy and spherically symmetric colloids[29].

To proceed we consider a colloid which we label 1. This colloid is bonded at $\Psi$ to patches on $n$ additional colloids. These $n$ colloids are labelled from 2 to $n + 1$. Now we integrate the position of the $n$ colloids over all possible bonding states of this cluster. We evaluate this integral with the mean value theorem as

$$\Xi^{(n)} = v_\Psi^n P^{(n)} \tag{34}$$

Where $P^{(n)}$ is the probability that $n$ colloids can be generated in the bond volume of $\Psi$ with no hard sphere overlap. The bond volume $v_\Psi$ is given as the surface area of the colloid which is devoid of patches integrated from the hard sphere diameter $d$ to the critical radius $r_c$

$$v_\Psi = \frac{r_c^3 - d^3}{3} \left( 4\pi - 2\pi \sum_{A \in \Gamma_p} (1 - \cos \theta_{c,A}) \right) \tag{35}$$

For the probability $P^{(n)}$ we employ the single cluster probabilities calculated in ref[29]. That is, we assume that $P^{(n)}$ can be approximated as the value obtained for the case that the center colloid 1 has no patches. We have not neglected the fact that the patches decrease the available surface area for patch-$\Psi$ attractions, this effect is included in the bond volume Eq. (35). This should be a reasonable approximation for competitive colloids since the maximum average number of bonds $\Psi$ could receive would be equal to the number of attractive patches. As patchy colloids typically have 1-5 patches, we should not expect strong competition for association at $\Psi$ as with mixtures of spherically symmetric and patchy colloids when the spherically symmetric colloids are dilute.[18,29,30]

### D) Competitive colloids with $\lambda$ equivalent patches

We now specialize the general approach developed in section B to the specific case of competitive colloids defined as having $\lambda$ equivalent monovalent patches $A$ with patch-patch attractions given by the energy scale $\varepsilon_{AA}$. In addition to patch-patch attractions, the $A$ patches are also attracted to the non-patch $\Psi$ region of the colloid with an energy scale $\varepsilon_{A\Psi}$. The ratio $R_{A\Psi}$ defined in Eq. (36) will then quantify the strength of $AA$ attractions in relation to $A\Psi$ attractions

$$R_{A\Psi} = \frac{\varepsilon_{AA}}{\varepsilon_{A\Psi}} \quad (36)$$

To solve for the bonding fractions $X_A$ we combine Eqns. (18), (20) and (22) – (23) to obtain the following closed equation for $X_A$

$$X_A + \lambda \xi \rho \kappa_{AA} f_{AA} X_A^2 + \frac{\rho X_A y_{HS}(d) \sum_{n=1}^{n_{\max}} \frac{n}{n!} \sqrt{\kappa_{AA}} f_{A\Psi} \Delta^{n-1} \Xi^{(n)} \delta^{(n)}}{1 + \sum_{n=1}^{n_{\max}} \frac{1}{n!} \Delta^n \delta^{(n)} \Xi^{(n)}} = 1 \quad (37)$$

with

$$\Delta = \lambda y_{HS}(d) \rho X_A f_{A\Psi} \sqrt{\kappa_{AA}} \quad (38)$$

Equation (37) can be interpreted as the sum over the bonding states of the $A$ patches. The first term on the left-hand side is the fraction of patches which are not bonded, the center term is the fraction of *A* patches which are bonded to other *A* patches $\chi_{AA}$, and finally the last term is the fraction of *A* patches which are bonded to the no patch $\Psi$ region $\chi_{A\Psi}$.

$$\chi_{AA} = \lambda \xi \rho \kappa_{AA} f_{AA} X_A^2;$$

(39)

$$\chi_{A\Psi} = \frac{\rho X_A y_{HS}(d) \sum_{n=1}^{n_{max}} \frac{n}{n!} \sqrt{\kappa_{AA}} f_{A\Psi} \Delta^{n-1} \Xi^{(n)} \delta^{(n)}}{1 + \sum_{n=1}^{n_{max}} \frac{1}{n!} \Delta^n \delta^{(n)} \Xi^{(n)}}$$

A quantity which describes the competition between *AA* and *A*$\Psi$ attractions is $\Theta$ which is defined as the ratio

$$\Theta = \frac{\chi_{AA}}{\chi_{A\Psi}}$$

(40)

For $\Theta > 1$ *AA* interactions dominate while for $\Theta < 1$ *A*$\Psi$ interactions dominate. Unlike Eq. (36), $\Theta$ depends on density and temperature. Upon solution of Eq. (37) for $X_A$ the fractions $X_\Psi$ can be solved for using Eqns. (18) and (21).

**III: Application to liquid water**

In this section, we apply the theory developed in II to develop a competitive colloid model which partially reproduces the anomalous properties of liquid water. The model is described by the number of patches $\lambda$, the size of the patches $\theta_c$, critical radius $r_c$, the patch-patch well depth $\varepsilon_{AA}$ and the patch-$\Psi$ well depth $\varepsilon_{A\Psi}$ as well as the colloidal diameter *d*. For the critical radius we assume a value $r_c = 1.1d$ which is consistent with previous experimental realizations[1,2] of patchy colloids. We bound the patch size at $\theta_c < 30°$ such that we can guarantee that the patches are singly bondable, hence satisfying the constraints of the theory. We do not define the colloidal diameter *d* as it will simply apply a scaling to our results. The parameters $\varepsilon_{AA}$, $\varepsilon_{A\Psi}$ and $\theta_c$ are then adjusted to minimize the error between the scaled saturated liquid densities of the colloid and the corresponding densities for water in the temperature range 0 °C < $T$ < 100 °C. We make comparisons in terms of scaled variables, due to the fact the diameter of water is on the order of angstroms and colloids would likely be on the order of nanometers to microns. We consider the following reduced densities

$$\rho^* = \rho d^3 \tag{41}$$

Neutron diffraction[43] shows that the first maximum in oxygen-oxygen correlation function in liquid water is located at a distance of 2.75 Å. SPC water[44] uses a diameter of 3 Å. In this work we determined a diameter of 2.85 Å gave the best agreement with experiment. Hence, when we compare to scaled water properties, we have scaled using a water diameter of $d_w$ = 2.85 Å.

We attempted patch numbers 2 ≤ λ ≤ 4, but the λ = 4 case is the only one for which we were able to reproduce the scaled densities of liquid water 0 °C < $T$ < 100 °C. Figure 4 compares scaled saturated liquid densities for water and our 4-patch competitive colloid. The patch parameters are given in table 1. Overall, the colloidal model is able to satisfactorily describe the saturated liquid densities of water over this temperature range.

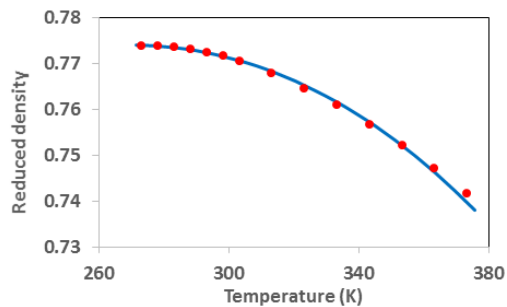

**Figure 4:** Competitive colloid model predictions (curve) and experimental data (symbols)[45] for the scaled saturated densities of liquid water

| Patches | $r_c$ | $\theta_c$ | $\varepsilon_{AA}/k_b$ (K) | $R_{A\Psi}$ |
|---|---|---|---|---|
| 4 | $1.1d$ | 22.24° | 3067.8 | 2.183 |

**Table 1:** Competitive colloid model parameters which reproduce the scaled liquid density of water at ambient conditions

In figure 5 we calculate the temperature-density phase diagram for the competitive colloid and compare to both saturated liquid densities of water in the temperature range 273 K $< T <$ 573 K and the liquid phase densities of super-cooled water in the temperature range 130 K $< T <$ 273K. Both the competitive colloid as well as liquid water show a decreasing density with decreasing $T$ in the super-cooled region. However, the competitive colloid has a monotonically decreasing density in this region. This is not the case for water which exhibits a density minimum near $T =$ 200 K. Finally, we note the critical temperature of the competitive colloid $T_c \sim$ 534 K is substantially lower than water $T_c =$ 647 K for water. This difference is a result of the fact that unlike water, the competitive colloids only exhibit short ranged attractions.

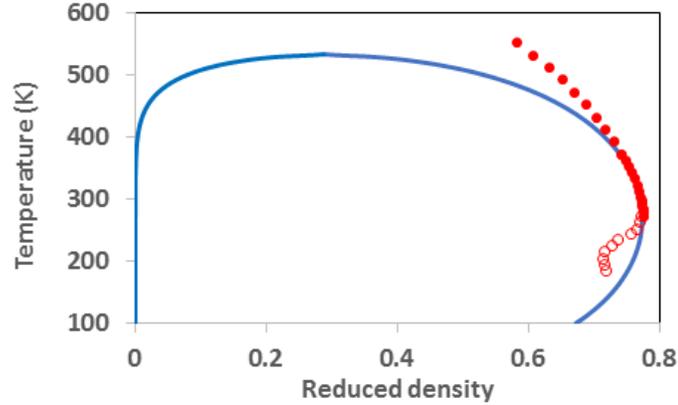

**Figure 5:** Model predictions (curve) of the phase diagram of competitive colloids compared to saturated liquid density data[45] (solid circles) and super-cooled density data[46] (open circles) of liquid water

Figure 6 shows the total average number of bonded patches per colloid, average number of patches participating in patch-$\Psi$ bonds per colloid, as well as the average number of patch-patch bonds per colloid. Starting from a high $T = 534$ K near the critical point, the number of patch-$\Psi$ bonds increases as temperature is decreased. Decreasing temperature results in an increase in the number of patch-patch bonds, eventually forcing a maximum in the number of patch-$\Psi$ bonds near $T = 315$ K. Figure 7 shows the distribution of $\chi_{\Psi,n}$ (Eq. 23) in the saturated liquid at $T = 300$ K. As can be seen, there is a non-negligible contribution from colloids which have $\Psi$ bonded up to 5 times. Further decrease in temperature results in an increase in the number of patch-patch bonds, and a decrease in patch-$\Psi$ bonds. It is this switching from patch-$\Psi$ bonds to patch-patch bonds which results in the density maximum.

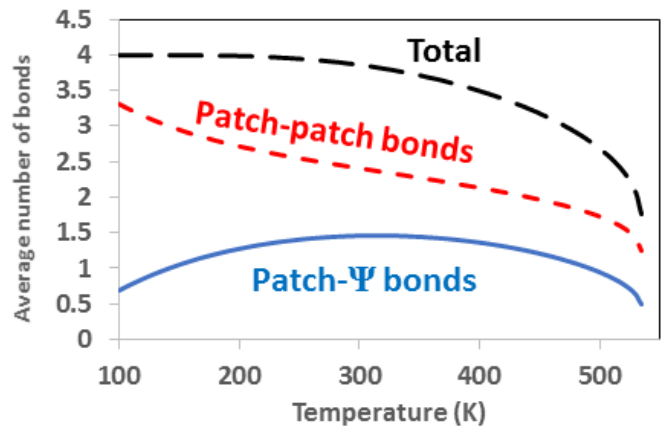

**Figure 6:** Competitive colloid model predictions for the average number of bonded patches per colloid (long dashed curve), number of patch-patch bonds per colloid (short dashed curve) and average number of patches participating in patch-$\Psi$ bonds per colloid (solid curve) for a saturated liquid

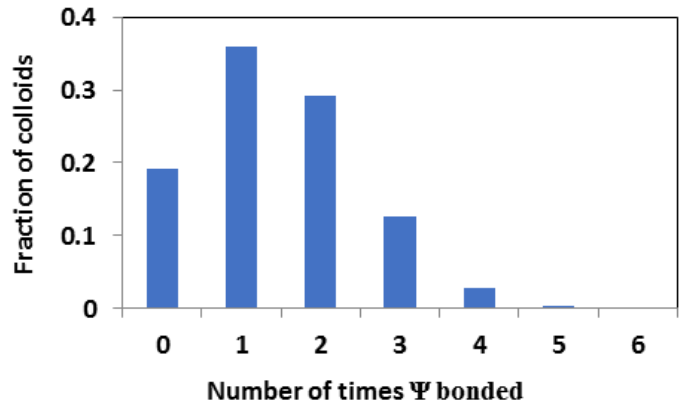

**Figure 7:** Fraction of colloids with $\Psi$ bonded $n$ times for a saturated liquid at $T = 300$ K

The transition of water towards a tetrahedral hydrogen bonding network results in a minimum in the isothermal compressibility $\kappa$ of liquid water.[47,48] In figure 8 we compare $\kappa^*$ =$\kappa$ /$\kappa$(303 K) versus $T$ for liquid phase and supercooled water to the competitive colloid model predictions for a saturated liquid. As can be seen, both the competitive colloid and water have their respective minima at $T$ = 319 K. The two scaled data sets are *in very good agreement for $T$ > 273 K. In the super-cooled water regime ($T$ < 273 K) the $\kappa$ of water increases more rapidly upon cooling than the competitive colloid. Like the density maximum, the minimum in $\kappa$ for the competitive colloid is a result of switching patch-$\Psi$ bonds to patch-patch bonds as temperature is lowered.

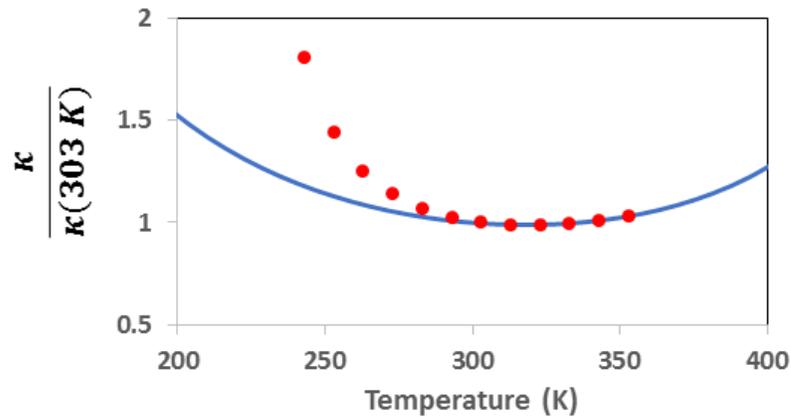

**Figure 8:** Comparison of competitive colloid model predictions for the scaled isothermal compressibility of a saturated liquid phase (curve) to experimental data[49] for water (symbols)

Finally, Fig. 9 compares the volume expansivity $\alpha$ = -$\partial$ln$\rho$/$\partial$T of liquid water at ambient conditions to the expansivity predicted for a saturated liquid by the competitive colloid model. Model and experiment are in good agreement for temperatures T > 270 K. The competitive colloid model successfully reproduces the anomalous feature of $\alpha$ < 0.

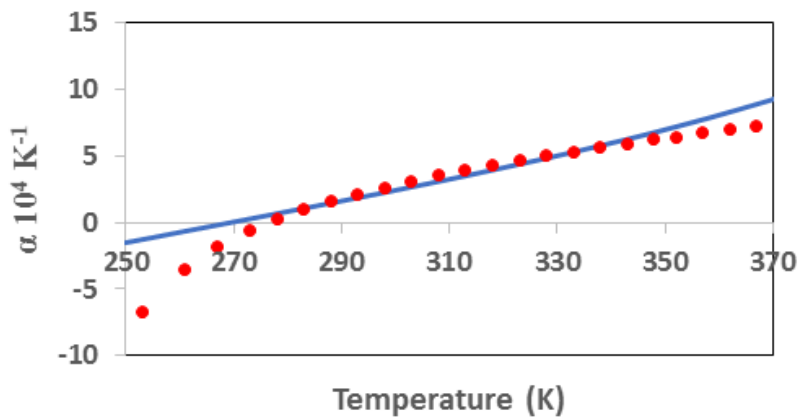

**Figure 9:** Comparison of experiment (symbols)[50] and competitive colloid predictions (curve) for the volume expansivity of liquid water.

## IV: Discussion and summary

We have developed a thermodynamic perturbation theory to describe the self-assembly and phase behavior of competitive colloids. As with previous models of patchy colloids[14,15,36], competition between different modes of attraction results in anomalous phase behavior. These previous patchy colloid models required two types of patches (bi-functional) to obtain competing energy scales. An advantage of competitive colloids is the plausibility of laboratory synthesis. It has been previously demonstrated that 4 patch colloids, with a single type of patch, can indeed be synthesized.[2]

As an application of the new theory, we adjusted the patch parameters to reproduce the density maximum of liquid water (appropriately scaled). Unlike water which has 2 association donor and 2 acceptor sites, we chose 4 equivalent patches for ease of laboratory synthesis. It was shown that the theory also predicted anomalous features of thermodynamic functions such as the minimum in the compressibility, consistent with liquid water. The anomalous properties of water are the result of the transition from normal fluid behavior at high temperatures to tetrahedral structure at ambient and super-cooled temperatures. The anomalous behavior of the competitive colloids is the result of the trading of patch-$\Psi$ bonds to patch-patch bonds as temperature is lowered.